\documentclass[sigconf,nonacm]{acmart}

\AtBeginDocument{%
  }

\setcopyright{acmlicensed}
\copyrightyear{2018}
\acmYear{2018}
\acmDOI{XXXXXXX.XXXXXXX}

\acmConference[Conference acronym 'XX]{Make sure to enter the correct
  conference title from your rights confirmation emai}{June 03--05,
  2018}{Woodstock, NY}

\acmISBN{978-1-4503-XXXX-X/18/06}

\usepackage[utf8]{inputenc}
\usepackage{amsmath} 
\usepackage{algorithm}
\usepackage{multirow}
\usepackage{algpseudocode} 
\usepackage{graphicx}
\usepackage{enumitem}

\usepackage{stfloats}

\begin{document}

\title{AdaptRec: A Self-Adaptive Framework for Sequential Recommendations with Large Language Models}


\author{Tong Zhang}
\authornote{Both authors contributed equally to this research.}
\email{Tong.Zhang-7@student.uts.edu.au}
\orcid{0000-0001-8043-237X}
\affiliation{%
  \institution{University of Technology Sydney}
  \city{Sydney}
  \state{NSW}
  \country{Australia}
}

\author{Dingxian Wang}
\authornotemark[1]
\email{Dingxian.Wang@student.uts.edu.au}
\affiliation{%
  \institution{University of Technology Sydney}
  \city{Sydney}
  \state{NSW}
  \country{Australia}
}

\author{Zihao Li}
\email{zihao.li@student.uts.edu.au}
\affiliation{%
  \institution{University of Technology Sydney}
  \city{Sydney}
  \state{NSW}
  \country{Australia}  
}

\author{Haoran Tang}
\email{haoran.tang@connect.polyu.hk}
\affiliation{%
  \institution{The Hong Kong Polytechnic University}
  \city{Hong Kong}
  \country{China}
}

\author{Jiancan Wu}
\authornote{Corresponding author.}
\email{wujcan@gmail.com}
\affiliation{%
  \institution{University of Science and Technology of China}
  \city{Hefei}
  \country{China}
}

\author{Xiang Wang}
\authornote{Corresponding author.}
\email{xiangwang@ustc.edu.cn}
\affiliation{%
  \institution{University of Science and Technology of China}
  \city{Hefei}
  \country{China}
}

\author{Guandong XU}
\authornote{Corresponding author.}
\email{gdxu@eduhk.hk}
\affiliation{%
  \institution{\begin{tabular}{@{}c@{}}The Education University of Hong Kong\end{tabular}}
  \city{Hong Kong}
  \country{China}
}

\renewcommand{\shortauthors}{Trovato et al.}




\received{20 February 2007}
\received[revised]{12 March 2009}
\received[accepted]{5 June 2009}


\begin{abstract}
The recent advancements in Large Language Models (LLMs) have generated considerable interest in their utilization for sequential recommendation tasks. While collaborative signals from similar users are central to recommendation modeling, effectively transforming these signals into a format that LLMs can understand and utilize remains challenging. The critical challenges include selecting relevant demonstrations from large-scale user interactions and ensuring their alignment with LLMs' reasoning process. To address these challenges, we introduce AdaptRec, a self-adaptive fram-ework that leverages LLMs for sequential recommendations by  incorporating explicit collaborative signals. AdaptRec employs a two-phase user selection mechanism --- User Similarity Retrieval and Self-Adaptive User Selection --- to efficiently identify relevant user sequences in large-scale datasets from multi-metric evaluation. We also develop a User-Based Similarity Retrieval Prompt, enabling the model to actively select similar users and continuously refine its selection criteria during training. Using the collaborative signals from similar users, we construct a User-Contextualized Recommendation Prompt that translates their behavior sequences into natural language, explicitly integrating this information into the recommendation process. Experiments demonstrate AdaptRec's superior performance, with significant improvements in HitRatio@1 scores of 7.13\%, 18.16\%, and 10.41\% across real-world datasets with full fine-tuning, and even higher gains of 23.00\%, 15.97\%, and 17.98\% in few-shot scenarios.
\end{abstract}

\maketitle

\section{Introduction}
Sequential recommendation systems are widely adopted to predict a user's next item of interest based on their historical interactions, with collaborative filtering (CF) serving as the cornerstone technique in these systems \cite{Quadrana2018SequenceAwareRS}. Recently, inspired by the remarkable abilities exhibited by Large Language Models \cite{brown2020language} --- including but not limited to the extensive world knowledge and in-context learning abilities --- researchers are increasingly exploring how to harness the potential of LLMs for advancing sequential recommendation (LLM4SRec). Scrutinizing existing studies on LLM4SRec, we can summarize a common paradigm that consists of two key steps:
(1) Converting the user's historical interaction sequence into input prompts in textual format;
(2) Performing recommendations using either tuning-based \cite{bao2023tallrec} or tuning-free LLMs \cite{liu2023chatgpt}.

%
%
%

\begin{figure*}[h]
    \centering
    \includegraphics[width=1\linewidth]{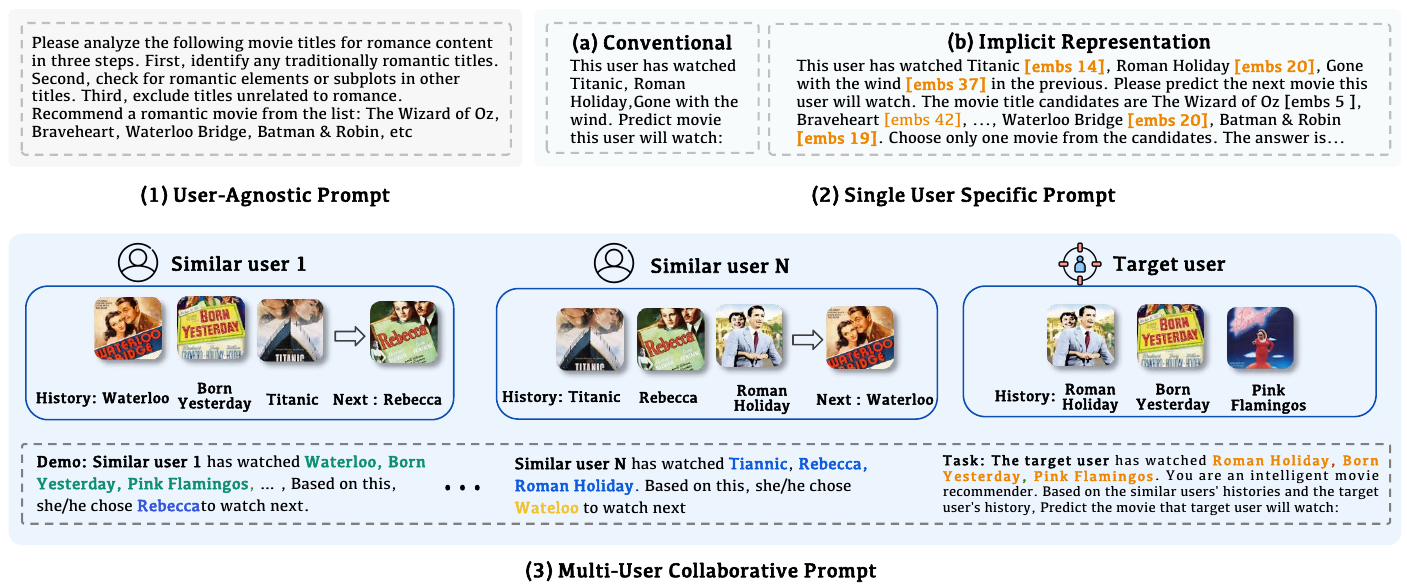}
    \vspace{-0.2cm}
    \caption{Overview of Prompt Design Strategies for Sequential Recommendation Systems: User-Agnostic Prompt, Single User-Specific Prompt, and Multi-User Collaborative Prompt.}
    \label{fig:comparison}
    \vspace{0.1cm}
\end{figure*}

Undoubtedly, as inputs to LLMs, prompts play a crucial role in guiding LLMs to generate outputs that align with task-specific requirements, as evidenced by advances across various domains \cite{wei2022chain,min2023recent,gao2020making,dong2022survey}. Recognizing this, researchers have explored diverse prompt design strategies, categorized into three main approaches based on how they incorporate user information, as illustrated in Figure 1: 
(1) User-Agnostic Prompts, these methods emphasis on guiding the recommendation process rather than user interaction \cite{wu2024coral, friedman2023leveraging, zhang2023collm}.
(2) Single User-specific Prompts focus on incorporating target user's interaction sequence, in either "isolated" or "implicit" way. By "isolated", we mean they focus exclusively on the sequential information of individual user \cite{liu2023chatgpt,chen2023knowledge,he2023large,shin2020autoprompt}, neglecting rich information from other users, while by "implicit", we mean they rely on learned embeddings or encoded signal \cite{bao2023tallrec, zheng2023boostinglarge, wang2023llmrec}.
(3) Multi-User Collaborative Prompts have emerged in recent studies, aiming to incorporate signals from similar users into the prompts \cite{wu2024coral,wang2023drdt}. 
These works select similar users through conventional collaborative filtering methods (e.g., Euclidean distance) and directly feed their sequences into prompts. However, such numerical similarity-based selection fails to transform collaborative signals into an explicit, interpretable format that aligns with LLMs' reasoning process.






To better bridge collaborative filtering and LLMs' reasoning capabilities, one promising solution is to leverage the in-context learning capabilities of LLMs, which enable them to adapt to new tasks using only a few demonstration examples in the prompt. This meth-od has shown impressive performance across a wide range of tasks \cite{brown2020language,dong2022survey}. In this light, we can explicitly introduce collaborative signals into prompts through providing informative demonstration examples, specifically other sequences exhibiting similar behavior patterns. While conceptually appealing, constructing high-quality prompts with demonstrations presents several challenges:

\begin{itemize}[leftmargin=2em]
    \item \textbf{Assessing Demonstration Quality:} The first challenge is determining whether a demonstration effectively conveys collaborative signals. As no standardized metrics exist for evaluating demonstration quality in sequential recommendation tasks, it is difficult to identify informative sequences.
    \item \textbf{Handling Large Search Spaces:} The second challenge is navigating the vast search space for relevant sequences. In large-scale datasets, this process is computationally intensive, requiring a balance between efficiency and accuracy to capture key collaborative signals.
    \item \textbf{Adapting Demonstration Selection:} The third challenge involves the static nature of existing methods for selecting similar users. This lack of adaptability results in poor
demonstration quality, leading to suboptimal recommendation performance.
\end{itemize}

\vspace{-0.2cm}
To address the aforementioned challenges, we propose a self-adaptive prompting framework that explicitly incorporates similar user's collaborative signals into sequential recommendation tasks. Our AdaptRec adaptively selects high-quality demonstration examples through a two-phase selection process, evaluates similar users using two distinct metrics from multiple perspectives to guarantee the quality and relevance of the selected demonstrations. \textbf{The framework consists of three core stages: (1) User Similarity Retrieval} employs collaborative filtering methods to identify potential similar users, effectively reducing the search space while retaining quality candidates for LLM selection; \textbf{(2) Self-Adaptive User Selection} introduces a User-based Similarity Retrieval Prompt, enabling LLMs to actively participate in selecting similar users, thus ensuring the demonstrations provide valuable collaborative signals. The model continuously updates its state during training, adapting its selection to align with its evolving understanding of the task; \textbf{(3) Contextual Prompt-based Recommendation} develops a User-Contextualized Recommendation Prompt. Here, similar users' behavior sequences are represented in human-readable language and used as demonstrations within the prompt. These demonstrations serve as explicit collaborative signals, guiding the model to generate more contextually relevant recommendations. In summary, our main contributions are:

\begin{itemize}[leftmargin=1em]
\item We introduce a novel, adaptive framework that explicitly integrates collaborative signals from similar users into sequential recommendation tasks. This approach leverages dynamic interaction data during training and inference, overcoming limitations of static user embeddings and enhancing performance in real-time environments.

\item We propose a two-phase selection strategy that enhances demonstration quality through complementary evaluation approaches: efficient collaborative filtering for initial retrieval and LLM-based assessment for quality refinement. To the best of our knowledge, this is the first work that enables LLMs to actively select collaborative signals rather than simply consuming pre-filtered demonstrations.

\item Extensive experiments demonstrate that our framework significantly outperforms state-of-the-art methods in both large-scale and few-shot scenarios. With fine-tuning, AdaptRec achieves improvements of 7.13\%, 18.16\%, and 10.41\% in HitRatio@1 across multiple real-world datasets. In few-shot scenarios, it shows improvements up to 23.00\%, 15.97\%, and 17.98\% compared to existing approaches.


\end{itemize}

\begin{figure*}[t]
    \centering
    \includegraphics[width=0.8\linewidth]{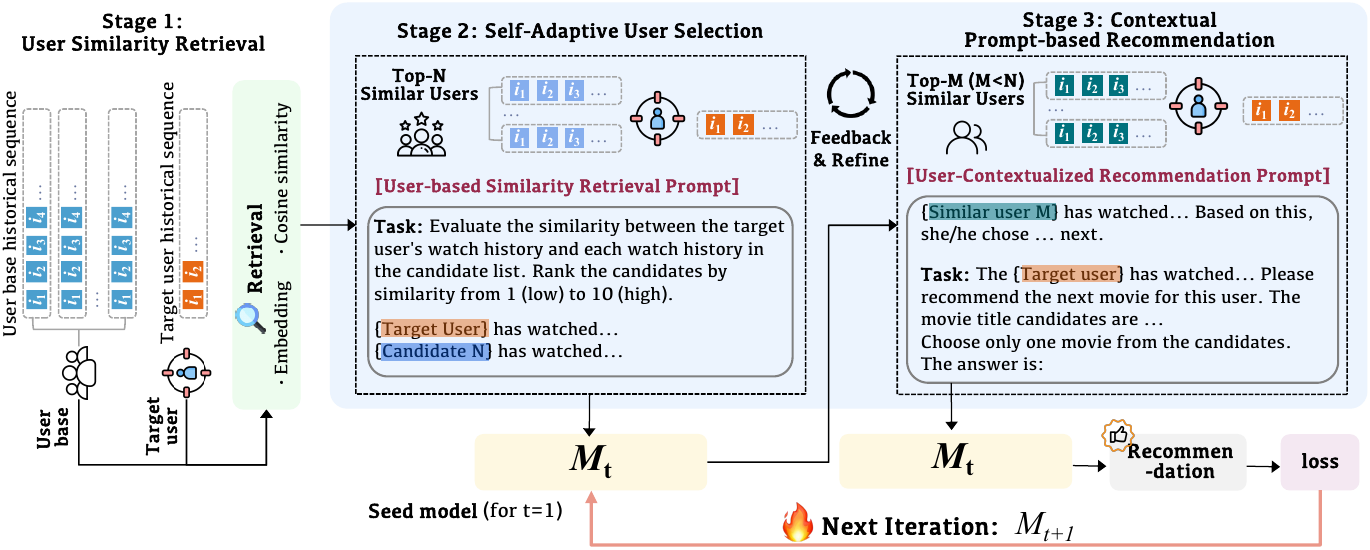}
    \vspace{-0.2cm}
    \caption{The Self-Adaptive User-Contextualized Sequential Recommendation Framework. (1) User Similarity Retrieval extracts relevant sequences. (2) Self-Adaptive User Selection refines similar user pool. (3) Contextual Prompt-based Recommendation generates personalized suggestions. An iterative feedback mechanism continuously improves user selection and recommendation accuracy.}
    \label{fig2}
    \vspace{-0.2cm}
\end{figure*}

\vspace{-0.2cm}

\section{RELATED WORK}

\subsection{Collaborative Information Modeling}

\textbf{Collaborative Filtering (CF)} stands as a milestone technique in recommendation systems, leveraging users' historical interactions to generate predictions. The core concept of CF involves recommending items by identifying users with similar behavioral patterns \cite{rendle2012bpr}. Early CF models primarily employed statistical techniques, such as user-item co-occurrence methods \cite{article}. A significant breakthrough came with the introduction of matrix factorization techniques \cite{hu2008collaborative,sarwar2001item}, which enhanced the capacity to capture latent factors underlying user preferences.
Prominent models like Matrix Factorization (MF) \cite{koren2009matrix} and Factored Item Similarity Model (FISM) \cite{kabbur2013fism} became influential in the field. Later,  neural network-enhanced models like AutoRec \cite{sedhain2015autorec} and Neural Matrix Factorization (NMF) \cite{he2017neural} moved beyond linear factorization, capturing more complex user-item interactions. Deep Matrix Factorization Models (DMF) \cite{xue2017deep} and DeepRec \cite{zhang2019deeprec} further demonstrated the power of deep architectures in improving recommendation accuracy. Despite its proven effectiveness and widespread adoption in traditional recommendation systems, the integration of collaborative information modeling techniques with Large Language Models (LLMs) in recommendation tasks remains an underexplored area, presenting a significant opportunity for advancement in the field of LLMRec.
These approaches have seen wide success across both academic research and industry applications, driving further exploration into collaborative information modeling for LLMRec. In this work, we propose leveraging collaborative information from similar users for LLMRec, a promising area that has yet to be explored.

\vspace{-1em}

\subsection{LLMs for Sequential Recommendation}
\textbf{LLMs} have emerged as powerful tools for sequential recommendation tasks, leveraging their superior language comprehension, reasoning capabilities, and world knowledge. The inherent similarity between sequential recommendation and next word prediction problems in NLP \cite{radford2018improving} has naturally led to the adoption of LLMs in this domain. Approaches utilizing LLMs can be divided into two paradigms based on whether parameters are tuned: non-tuning and tuning. The non-tuning paradigms involves directly prompting LLMs  to generate recommendations \cite{he2023large,hou2024large,liu2023chatgpt,zhang2023chatgpt}, leveraging their general reasoning and semantic abilities. However, this approach often struggles with the misalignment between general language tasks and personalized recommendation data. In contrast, the tuning paradigm employs prompt learning or instruction tuning  to better align LLMs with recommendation tasks\cite{bao2023bi,bao2023tallrec,wu2024exploring,zhang2023recommendation}, enabling them to capture user-item interactions more effectively. 

Recent efforts to enhance LLMRec have focused on integrating collaborative information. For example, Qiu et al. \cite{qiu2021u} developed U-BERT, which utilizes user reviews to learn user representations. Hua et al. \cite{hua2023index} investigated ID encoding through vocabulary expansion, effectively incorporating user and item IDs into the language model's token space. Hou et al. \cite{hou2022towards} proposed a BERT-based framework that incorporates item descriptions, enhancing the model's understanding of item characteristics. While these approaches make progress in utilizing collaborative data, they primarily rely on static information or individual user sequences, overlooking the rich collaborative signals from similar users' behavioral patterns that have proven effective in traditional recommender systems. This limitation is particularly significant in sequential recommendations, where user preferences evolve over time. Moreover, due to the token limitations of large language models, even with prompt tuning, the context within the prompt remains constrained, making the selection of the most essential collaborative information a critical research challenge. To address these gaps, we propose a comprehensive framework that combines user similarity retrieval, self-adaptive user selection, and contextual prompt-based recommendation. Our approach includes an iterative feedback mechanism that continuously refines both the similar user selection and the recommendation generation, leading to more accurate and personalized sequential recommendations.

\section{PRELIMINARY}
\sloppy
\subsection{Problem Formulation}
In sequential recommendation, we are given a user set $\mathcal{U}$ and an item set $\mathcal{I}$. Each user $v \in \mathcal{U}$ has chronologically engaged with item sequence  $S_v = [i_1, i_2, ..., i_n]$, where $i_j \in \mathcal{I}$ and $n$ denotes the sequence length. Based on the historical interactions, sequential recommender systems are used to predict the next item $i_{n+1}$ that the user will interact with. Additionally, we incorporate collaborative signals from similar users $\mathcal{U}_2 \subseteq \mathcal{U}$, with their corresponding interaction histories $H = \{ H_1, H_2, \dots, H_n \}$, to enhance recommendation accuracy.

\subsection{Parameter-Efficient Fine-Tuning}
Given that LLMs typically consist of billions of parameters, full tuning is a resource-intensive and time-consuming process. In this work, we employ \textbf{LoRA} \cite{hu2022lora}, a prominent and widely adopted Parameter-Efficient Fine-Tuning (PEFT) solution \cite{houlsby2019parameter}. To make fine-tuning
more efficient, LoRA adds pairs of rank-decomposition weight matrices to the existing weights of the LLM in a modular manner. Crucially, during fine-tuning, only these low-rank matrices are trained, while the original weights remain frozen. The training objective of LoRA can be formulated as follows:
\begin{equation}
\begin{aligned}
   \max_{\Theta} \sum_{v \in \mathcal{U}} \sum_{t=1}^{|i_{n+1}|} & \log P_{W_{\text{pt}}+AB}
\end{aligned}
\end{equation}

where $P_{W_{\text{pt}}+AB}$ denotes the model's probability distribution, $W_{\text{pt}} \in \mathbb{R}^{d \times d}$ represents the frozen pre-trained weights, and $A \in \mathbb{R}^{d \times r}, B \in \mathbb{R}^{r \times d}$ are trainable low-rank decomposition matrices with $r \ll d$, making the number of trainable parameters significantly smaller compared to $W_{\text{pt}}$.

\section{METHODOLOGY}
In this section, we present a comprehensive overview of AdaptRec, detailing its framework architecture and learning process. We then delve into the key components of our approach.

\subsection{AdaptRec Framework}
To effectively utilize explicit collaborative signals from similar users, we propose a self-adaptive prompting framework, as illustrated in Figure \ref{fig2}. Our framework consists of three main stages: (1) User Similarity Retrieval: This initial stage performs coarse-grained filtering of user sequences, effectively narrowing down the vast search space of user interactions. (2) Self-Adaptive User Selection: This phase utilizes a User-based Similarity Retrieval Prompt, enabling the model to actively select similar users and continuously refine its selection criteria during training. (3) Contextual Prompt-based Recommendation: we construct a User-Contextualized Recommendation Prompt, translate similar user's behavior sequences into natural language, explicitly integrating this information into the recommendation process. 

\subsection{User Similarity Retrieval}
Given the token length limitations of LLMs, directly retrieving similar users across the entire dataset is computationally impractical. Thus, in this initial stage, we employ a coarse-grained filtering process to identify relevant candidates. The similarity between users is quantified using the cosine similarity of their respective item sequence title embeddings. For a target user $v$ and a user $u \in \mathcal{U}$, the sequence embeddings are denoted as $\mathbf{e}_v$ and $\mathbf{e}_u \in \mathbb{R}^d$, respectively. The cosine similarity between these embeddings is computed as:
\begin{equation}
\text{sim}(v, u) = \frac{\mathbf{e}_v \cdot \mathbf{e}_u}{\|\mathbf{e}_v\| \|\mathbf{e}_u\|},
\end{equation}

\noindent where $\mathbf{e}_v \cdot \mathbf{e}_u$ represents the dot product of the embeddings, and $\|\mathbf{e}_v\|$ and $\|\mathbf{e}_u\|$ are the corresponding Euclidean norms. Based on this similarity measure, we select the top-$N$ most similar users to form the initial subset:
\begin{equation}
\mathcal{U}_1 = \{ u \mid u \in \mathcal{U}, \text{rank}(\text{sim}(v, u)) \leq N \},
\end{equation}

\noindent where $\mathcal{U}_1$ represents the set of top-$N$ similar users for the target user $v$, ranked according to the cosine similarity scores.

\subsection{Self-Adaptive User Selection}
\begin{figure}[h]
    \includegraphics[width=0.48\textwidth]{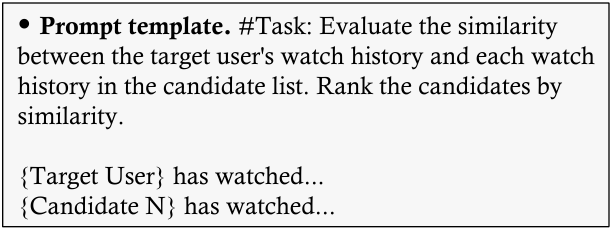}
    \caption{User-Based Similarity Retrieval Prompt}
    \label{fig:03}
\end{figure}

Based on the top-$N$ similar users subset $\mathcal{U}_1$ obtained from coarse-grained retrieval phase, we develop a User-Based Similarity Retrieval Prompt, as illustrated in Figure \ref{fig:03}, to let LLMs select the most informative demonstration examples for the subsequent recommendation stage.

Through this prompt template, for each user $u \in \mathcal{U}_1$, the model evaluates their sequence similarity with the target user $v$ by examining their interaction patterns $[i_1, i_2, ..., i_n]$. This evaluation process can be formalized as:
\begin{equation}
R(v, u) = f(\text{sim}(v, u), \mathbf{h}_v, \mathbf{h}_u),
\end{equation}

\noindent where $\text{sim}(v, u)$ is the initial cosine similarity from the previous stage, and $\mathbf{h}_v, \mathbf{h}_u \in \mathbb{R}^d$ are the hidden state representations capturing the contextual patterns in users' interaction sequences. The function $f$ represents the LLM's similarity assessment process guided by our designed prompt template. Based on these relevance scores, LLMs select the most similar users to form the refined subset:
\begin{equation}
\mathcal{U}_2 = \{ u \mid u \in \mathcal{U}_1, \text{rank}(R(v, u)) \leq M \},
\end{equation}

\noindent where $M < N$. Although the selection process itself does not involve model training, the selected demonstrations directly influence the subsequent recommendation stage where the model is fine-tuned. This iterative mechanism enables the model to progressively refine its selection criteria as its recommendation capability evolves through the training process.

\subsection{Contextual Prompt-based Recommendation}
\begin{figure}[h]
    \vspace{-10pt}
    \includegraphics[width=0.48\textwidth]{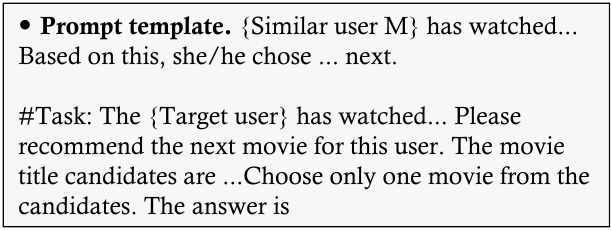}
    \vspace{-15pt}
    \caption{Contextual Prompt-based
Recommendation}
    \label{fig:04}
\end{figure}
In last phase, top-$M$ similar users are incorporated as demonstrations in the prompt, as demonstrated in Figure \ref{fig:04} to enhance the recommendation process. For the target user $v$ with interaction sequence $[i_1, i_2, ..., i_n]$, we formulate the probability of the next item as:
\begin{equation}
P(i_{n+1,t} \mid [i_1, i_2, ..., i_n], \{H_u\}_{u \in \mathcal{U}_2}, i_{<t}),
\end{equation}

\noindent where $i_{n+1,t}$ refers to the $t$-th token of the target item $i_{n+1}$, and $i_{n+1,<t}$ indicates the tokens preceding $i_{n+1,t}$. This formulation leverages both the target user's historical interactions and the collaborative signals from similar users in $\mathcal{U}_2$.

\subsection{Model Learning}

The optimization of our sequential recommender is conducted by minimizing the negative log-likelihood loss, a method commonly employed in LLM-based recommendation systems. The model is fine-tuned using the contextual prompts that incorporate the target user's interaction history and the demonstrated sequences from the top-$M$ similar users.
\subsubsection{Training Objective}
For a target user $v$ with an interaction sequence $S_v = [i_1, i_2, ..., i_n]$, the training objective is to minimize the following loss:
\begin{equation}
\mathcal{L} = - \sum_{v \in \mathcal{U}} \log P(i_{n+1} \mid S_v, \{H_u\}_{u \in \mathcal{U}_2}).
\end{equation}
\noindent Here, $P(i_{n+1} \mid S_v, \{H_u\}_{u \in \mathcal{U}_2})$ represents the conditional probability of recommending the correct next item, given the target user's sequence $S_v$ and the sequences from selected similar users in $\mathcal{U}_2$.

\subsubsection{Training Process}
During training, the model performs the recommendation task based on the contextual prompts. The negative log-likelihood loss is computed at each iteration, and the model parameters are updated accordingly. The updates, including those applied to the LoRA-augmented parameters, follow the standard gradient descent process:
\begin{equation}
W_{t+1} = W_t - \eta \nabla \mathcal{L}(W_{\text{pt}} + AB),
\end{equation}

\noindent where $W_t$ represents the model parameters at iteration $t$, $\eta$ is the learning rate, and $AB$ denotes the product of trainable low-rank matrices introduced by LoRA. The training process continues iteratively until the model converges or reaches a predefined number of training steps.

\section{EXPERIMENTS AND RESULTS}
In this section, we evaluate AdaptRec across three real-world datasets against both traditional sequential recommenders and state-of-the-art LLM-based approaches. To demonstrate the effectiveness of our framework, we carry out extensive experiments to answer the following main research questions:

\begin{itemize}[leftmargin=1em]
    \item \textbf{RQ1}: How does AdaptRec perform compared to traditional sequential models and LLM-based recommendation approaches?
     \item \textbf{RQ2}: How effective is our coarse-grained retrieval compared to random sampling in identifying relevant users?
     \item \textbf{RQ3}: How does the self-adaptive user selection process compare to static demonstration selection in improving recommendation accuracy?
    \item \textbf{RQ4}: What is the impact of user-based contextual prompts on the recommendation, and how does the number of demonstrations affect model performance?

\end{itemize}

\begin{table}[t]
    \centering
    \caption{Statistics of Datasets}
    \begin{tabular}{cccc}
    \hline
    \textbf{Dataset} & \textbf{MovieLens} & \textbf{LastFM} & \textbf{GoodReads} \\
    \hline
    Sequence& 943 & 1,220 & 1,120 \\
    Item & 1,682 & 4,606 & 2,359 \\
    Interaction & 100,000 & 73,510 & 73,637 \\
    \hline
    \end{tabular}
    \label{tab:my_label}
\end{table}

\subsection{Experimental Settings}
\subsubsection{Datasets}
\begin{itemize}[leftmargin=1em]
\item \textbf{MovieLens}\cite{harper2015movielens} is a commonly-used movie recommendation dataset that contains user ratings and movie titles.
    
\item \textbf{LastFM} \cite{cantador2011second} collected from the Last.fm online music platform, includes user-artist listening relationships and the names of artists
    
\item \textbf{GoodReads}    \cite{DBLP:conf/recsys/WanM18} is a large-scale book recommendation dataset that includes user ratings, reviews, and detailed book metadata.
\end{itemize}
Considering the significant time required to fine-tune LLMs compared to traditional recommendation systems, we selected the MovieLens100K dataset to ensure a manageable experiment size. For the GoodReads dataset, we focused on the "History" genre, applying stringent filters: removing users with fewer than 40 reviews and excluding entries with missing book titles in the metadata, preserving their interactions to form a moderately sized dataset. To ensure proper temporal alignment and avoid information leakage, interactions were arranged chronologically, and the data was split into training, validation, and test subsets in an 8:1:1 ratio. Detailed statistics of the datasets are provided in Table \ref{tab:my_label}.

\begin{table*}[h]
    \centering
    \renewcommand{\arraystretch}{1.2}
    \caption{\centering Performance comparison of different methods. The best results are in bold and the second best results are underlined. The row “Improv.” indicates the relative performance gain of our AdaptRec and the suboptimal method.}
    \renewcommand{\arraystretch}{1.1}
    \begin{tabular}{llccccccccc}
        \toprule
        & & \multicolumn{3}{c}{\textbf{Traditional Methods}} & \multicolumn{4}{c}{\textbf{LLM-based Methods}} & & \\
        \cmidrule(lr){3-5} \cmidrule(lr){6-9}
        Dataset & Metrics & GRU4Rec & Caser & SASRec & Llama2 & GPT-4 & MoRec & LLaRA & AdaptRec & Improv.\\ \midrule
        \multirow{4}{*}{MovieLens} 
        & HR@1 & 0.3750 & 0.3861 & 0.3444 & 0.0421 & 0.2000 & 0.2822 & \underline{0.4421} & \textbf{0.4736} & +7.13\% \\[1.5pt]
        & NDCG@5 & 0.5625 & 0.5715 & 0.5416 & 0.3168 & 0.4500 & 0.5028 & \underline{0.6105} & \textbf{0.6527} & +6.91\% \\[1.5pt]
        & NDCG@20 & 0.5875 & 0.5972 & 0.5652 & 0.3335 & 0.4700 & 0.5246 & \underline{0.6379} & \textbf{0.6818} & +6.88\% \\[1.5pt]
        & ValidRatio & 1.0000 & 1.0000 & 1.0000 & 0.4421 & \underline{0.9895} & \textbf{1.0000} & 0.9684 & 0.9684 & -3.16\% \\ \midrule
        \multirow{4}{*}{LastFM}
        & HR@1 & 0.2616 & 0.2233 & 0.2233 & 0.0246 & 0.3770 & 0.1652 & \underline{0.4508} & \textbf{0.5327} & +18.16\% \\[1.5pt]
        & NDCG@5 & 0.4904 & 0.4650 & 0.4650 & 0.3049 & 0.5639 & 0.4239 & \underline{0.6159} & \textbf{0.6663} & +8.18\% \\[1.5pt]
        & NDCG@20 & 0.5116 & 0.4853 & 0.4853 & 0.3210 & 0.5889 & 0.4424 & \underline{0.6435} & \textbf{0.6960} & +8.16\% \\[1.5pt]
        & ValidRatio & 1.0000 & 1.0000 & 1.0000 & 0.3443 & \textbf{1.0000} & \textbf{1.0000} & \underline{0.9918} & 0.9934 & -0.66\% \\ \midrule
        \multirow{4}{*}{GoodReads}
        & HR@1 & 0.3512 & 0.3348 & 0.3024 & 0.0210 & 0.3104 & 0.1438 & \underline{0.4014} & \textbf{0.4432} & +10.41\% \\[1.5pt]
        & NDCG@5 & 0.5459 & 0.5348 & 0.5134 & 0.3026 & 0.5187 & 0.4098 & \underline{0.5804} & \textbf{0.6112} & +5.31\% \\[1.5pt]
        & NDCG@20 & 0.5698 & 0.5581 & 0.5358 & 0.3186 & 0.5414 & 0.4277 & \underline{0.6064} & \textbf{0.6387} & +5.33\% \\[1.5pt]
        & ValidRatio & 1.0000 & 1.0000 & 1.0000 & 0.3102 & \textbf{1.0000} & \textbf{1.0000} & \underline{0.9234} & \textbf{1.0000} & 0.00\% \\ \bottomrule
    \end{tabular}
    \label{tab:performance_comparison}
\end{table*}

\subsubsection{Baselines}
\label{subsec:baselines}
\begin{itemize}[leftmargin=1em]
    \item \textbf{Traditional Sequential Recommenders:} GRU4Rec \cite{hidasi2015session}, Caser \cite{tang2018personalized}, and SASRec \cite{kang2018self}, are RNN-based, CNN-based, and attention-based sequential recommenders, respectively.
    \item \textbf{LLM-based Models:} (1) Llama \cite{touvron2023llama} is a well-known open-source LLM released by Meta. (2) GPT-4 \cite{openai2024gpt4technicalreport}, released by OpenAI, is a milestone of LLMs excelling in various tasks. (3) MoRec \cite{yuan2023go} enhances the traditional recommenders by encoding item's modality features, such as text features. (4) LLaRA \cite{liao2024llara} develops a hybrid embedding that incorporates both item textual information and behavioral signals for recommendation.
\end{itemize}

\subsubsection{\textbf{Implementation Details}}
We adopt Llama-2-7B \cite{touvron2023llama} as our base LLM, employing dynamic instruction sampling to enhance interface flexibility. Traditional recommenders are implemented following \cite{yang2023generic}, utilizing Adam optimization (lr=1e-3, d=64,
 batch=256) with L2 regularization optimized via grid search. LLM-based methods are trained for 5 epochs (batch=128) with learning rate warm-up and cosine scheduling. Results are averaged over five runs to ensure statistical significance. 
\subsubsection{\textbf{Evaluation Metrics}}
Following the leave-one-out strategy, we evaluate each model on a candidate set of 20 items (one ground truth and 19 negative samples). We employ Hit Ratio (HR@1) to assess recommendation accuracy and Normalized Discounted Cumulative Gain (NDCG@5/20) to measure ranking quality. Additionally, we propose a new metric---valid ratio---for LLM-based models to quantify the proportion of valid recommendations within the candidate set, as LLMs may generate outputs beyond the predefined candidates during prompting. As for traditional models that select candidates based on predicted probabilities, we regard their valid ratio as 1. This multi-faceted evaluation framework enables rigorous comparison of recommendation accuracy and adherence to task constraints across diverse model architectures.

\begin{table*}[h]
  \centering
  \caption{The results of ablation study on AdaptRec}
  \label{tab:ablation}
  \renewcommand{\arraystretch}{1.1}
  \begin{tabular}{l ccc ccc ccc}
      \toprule
      \multirow{2}{*}{Method} & \multicolumn{3}{c}{MovieLens} & \multicolumn{3}{c}{LastFM} & \multicolumn{3}{c}{GoodReads} \\
      \cmidrule(lr){2-4} \cmidrule(lr){5-7} \cmidrule(lr){8-10}
      & HR@1 & NDCG@5 & NDCG@20 & HR@1 & NDCG@5 & NDCG@20 & HR@1 & NDCG@5 & NDCG@20 \\
      \midrule
      w/o retrieval & 0.3224 & 0.4826 & 0.5012 & 0.3315 & 0.4923 & 0.5124 & 0.3612 & 0.5124 & 0.5358 \\
      w/o self-adaptive & 0.4521 & 0.6127 & 0.6382 & 0.5127 & 0.6358 & 0.6624 & 0.4353 & 0.5912 & 0.6187 \\
      w/o demo & 0.3512 & 0.5134 & 0.5358 & 0.3678 & 0.5259 & 0.5491 & 0.4432 & 0.5824 & 0.6081 \\
       AdaptRec & \textbf{0.4736} & \textbf{0.6527} & \textbf{0.6818} & \textbf{0.5327} & \textbf{0.6663} & \textbf{0.6960} & \textbf{0.4432} & \textbf{0.6112} & \textbf{0.6387} \\
      \bottomrule
  \end{tabular}
\end{table*}

\subsection{Performance Comparison (RQ1)}
We compare our proposal with conventional sequential recommenders and LLM-enhanced sequential recommenders, as detailed in Section~\ref{subsec:baselines}. The results are reported in Table~\ref{tab:performance_comparison} from which we observe:

\begin{itemize}[leftmargin=1em]
   \item \textbf{AdaptRec outperforms baselines:} AdaptRec demonstrates superior performance across all evaluation metrics on three datasets. Specifically, it achieves the highest HitRatio@1 scores of 0.4736, 0.5327, and 0.4432 on MovieLens, LastFM, and GoodReads respectively, surpassing the next best method by 7.13\%, 18.16\%, and 10.41\%. The improvement extends to ranking quality, with highest NDCG@5 and NDCG@20 on the three datasets. This consistent outperformance across different metrics and datasets underscores the efficacy of our self-adaptive framework in combining collaborative signals with LLM capabilities for sequential recommendation.
   
   \item \textbf{Traditional sequential recommenders:} Caser, and SASRec exhibit consistently lower performance across all datasets. Even the best traditional model only achieves NDCG@20 of 0.5972 on MovieLens and 0.5581 on GoodReads, falling short by a significant margin. This performance gap can be attributed to these models' reliance solely on behavioral patterns, while AdaptRec benefits from integrating semantic understanding through LLMs. The results demonstrate that conventional sequential patterns, though fundamental, are insufficient for capturing the complex user-item relationships that can be understood through language models.
   
   \item \textbf{LLM-based method results:} a) Vanilla LLMs (Llama2 and GPT-4) show evident limitations on recommendation task. While GPT-4 maintains a high valid ratio (0.98) in generating recommendations within candidate sets, Llama2's valid ratio drops significantly (0.31-0.44), revealing its instability in controlled generation. b) Specialized LLM recommenders (MoRec and LLaRA) demonstrate improved stability but still underperform compared to AdaptRec. This comparison highlights that neither direct application of LLMs nor simple LLM enhancement is sufficient - the key advantage of AdaptRec lies in effectively combining LLM's semantic understanding with sequential patterns through its self-adaptive framework.
   \item Dataset characteristics significantly influence model performance. The most substantial improvement (18.16\%) is observed on LastFM, where item descriptions (artist names) are predominantly in English, aligning well with LLMs' training data. In contrast, the relatively smaller improvement on MovieLens (7.13 \%) can be attributed to its multilingual movie titles, which may challenge LLMs' language understanding capabilities. This pattern reveals that the effectiveness of LLM-based recommendation methods is closely tied to the language composition of item descriptions, suggesting potential directions for enhancing multilingual recommendation scenarios.
   \item  Despite its superior performance, we identify several key limitations: 
a) The model shows reduced effectiveness with multilingual content, as evidenced by the smaller improvement on MovieLens (7.13\%) compared to LastFM (18.16\%). 
b) The iterative nature of our self-adaptive framework introduces additional computational overhead compared to traditional approaches, though partially mitigated through LoRA adaptation. 
c) While achieving high valid ratio (0.96), AdaptRec still falls short of the perfect validity demonstrated by traditional models, indicating room for improvement in controlled generation.
\end{itemize}

\begin{figure*}[htbp] 
    \includegraphics[width=0.95\textwidth]{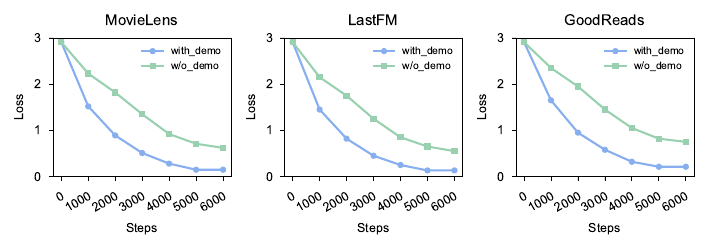} 
    \vspace{-15pt} 
    \caption{Training Loss Trends Across Training Steps}
    \label{fig6}
\end{figure*}


\subsubsection{\textbf{Model Efficiency Analysis}}
The efficiency of recommendation models is crucial for practical deployment. We analyze the computational efficiency by examining the training convergence patterns across different datasets. As illustrated in Figure~\ref{fig6}, AdaptRec demonstrates superior convergence efficiency compared to the baseline without demonstrations. Specifically, AdaptRec achieves convergence within 5,000 training steps, with loss values stabilizing at 0.148, 0.132, and 0.212 for MovieLens, LastFM, and GoodReads, respectively. In contrast, the baseline model exhibits slower convergence, with loss values at 6,000 steps remaining at 0.62, 0.55, and 0.75 for the respective datasets, indicating non-convergence even after extended training. The accelerated convergence stems from the effective guidance of demonstrations, enabling more efficient learning of user preferences. This efficiency advantage, combined with superior recommendation accuracy, establishes AdaptRec as a practically viable solution for real-world recommendation scenarios.

\begin{figure}[t]
    \includegraphics[width=0.48\textwidth]{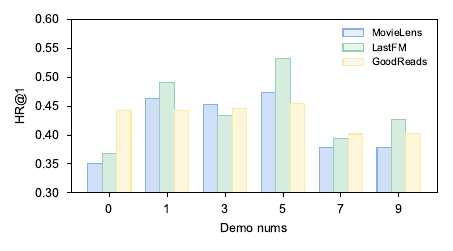}
    \vspace{-15pt}
    \caption{Performance with Different Numbers of Demonstrations across three datasets}
    \label{fig5}
\end{figure}

\begin{table}[t]
  \centering
  \caption{Impact of varying demonstration numbers on HR@1 across three datasets}
  \label{tab:hr_impact}
  \setlength{\tabcolsep}{6pt}
  \begin{tabular}{c ccc}
      \toprule
      
      Demos (M) & MovieLens & LastFM & GoodReads \\
      \midrule
      M=1 & 0.4631 & 0.4912 & 0.4417 \\
      M=3 & 0.4526 & 0.4344 & 0.4464 \\
      M=5 & \textbf{0.4736} & \textbf{0.5327} & \textbf{0.4553} \\
      M=7 & 0.3789 & 0.3934 & 0.4017 \\
      M=9 & 0.3789 & 0.4262 & 0.4021 \\
      \bottomrule
  \end{tabular}
\end{table}

\begin{figure}[t]
   \centering
   \includegraphics[width=0.78\linewidth]{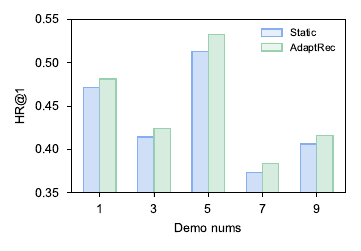}
   \caption{Performance comparison of static and adaptive selection strategies on MovieLens dataset}
   \label{fig:selection_comparison}
\end{figure}

\subsection{Effectiveness of User Similarity Retrieval (RQ2)}
To evaluate our coarse-grained filtering mechanism, we compare:
\begin{itemize}[leftmargin=1em]
   \item \textbf{Random Selection}: Randomly samples users from the candidate pool (w/o retrieval)
   \item \textbf{Similarity Retrieval}: Our proposed coarse-grained filtering approach
\end{itemize}

We conduct experiments with 5 demonstrations across datasets. Table~\ref{tab:ablation} reveals three key findings: (1) Random selection significantly underperforms our retrieval approach, with substantial drops in HR@1 - decreasing by 31.93\% (0.4736 to 0.3224) on MovieLens, 37.77\% (0.5327 to 0.3315) on LastFM, and 18.50\% (0.4432 to 0.3612) on GoodReads.(2) This performance gap extends across evaluation metrics, with similar degradation patterns observed in NDCG@5 and NDCG@20, indicating that the impact of effective user selection persists across different recommendation scenarios.(3) The consistent underperformance of random selection highlights that simply increasing the candidate pool size without proper filtering fails to identify behaviorally relevant users. This validates the necessity of our coarse-grained retrieval mechanism for maintaining recommendation quality.

\subsection{Impact of Self-Adaptive Similar User Selection (RQ3)
}
To verify the contribution of self-adaptive user selection mechanism, we conduct comparison experiments with two variants: 
\begin{itemize}[leftmargin=1em]
    \item \textbf{Static Selection}: Randomly samples similar users from the retrieved candidate pool as demonstrations.
    \item \textbf{Self-Adaptive Selection}: Our proposed approach that employs self-adaptive selection after the User Similarity Retrieval stage.
\end{itemize}
We conduct experiments with different demonstration numbers (N=1,3,5,7,9) across three datasets. Figure~\ref{fig:selection_comparison} illustrates the comparative results:
First, Self-Adaptive Selection consistently outperforms static selection across all settings, achieving HR@1 improvements of 3.21\%, 3.45\%, and 3.32\% on MovieLens, LastFM and GoodReads respectively at N=5. This demonstrates the advantage of dynamically selecting similar users based on behavioral patterns rather than random assignment. More importantly, this performance gain persists even with limited demonstration options - at N=1 and N=3, Self-Adaptive Selection maintains an average HR@1 improvement of 2.31\% across datasets. This indicates our mechanism's robustness in extracting meaningful collaborative signals even from small candidate pools. Such consistent performance confirms highlighting the value of incorporating LLM's reasoning capabilities in demonstration selection.

\subsection{Impact of User-Based Contextual Prompt (RQ4)}

To understand how contextual information influences recommendation quality, we compare:
\begin{itemize}[leftmargin=1em]
   \item \textbf{Baseline}: Standard recommendation without demonstrations (w/o demo)
   \item \textbf{Contextual Prompting}: Our approach with varying demonstration numbers (N=1,3,5,7,9)
\end{itemize}

Table~\ref{tab:hr_impact} and Figure~\ref{fig5} present the results, revealing several key findings: (1) The introduction of contextual demonstrations brings substantial performance gains. With just one demonstration, the model achieves significant improvements in HR@1 across all datasets - from 0.3512 to 0.4631 on MovieLens and from 0.3678 to 0.4912 on LastFM, corresponding to relative gains of 31.86\% and 33.55\%. Similar trends are observed in NDCG@5, with improvements of 16.34\% and 17.59\% respectively. These consistent gains demonstrate that even minimal contextual information enables the model to better capture user preferences. (2) The performance exhibits an inverted U-shaped pattern, reaching optimal results at N=5 across all datasets. At this sweet spot, the model achieves its highest HR@1 scores of 0.4736 (MovieLens), 0.5327 (LastFM), and 0.4553 (GoodReads). The corresponding NDCG@5 scores peak at 0.6527, 0.6663, and 0.6112, suggesting that five demonstrations provide an ideal balance of contextual information. (3) Beyond this optimal point, increasing demonstrations leads to notable performance degradation. When N increases from 5 to 7, HR@1 drops sharply by 26.15\% on LastFM and 20.00\% on MovieLens. This consistent decline indicates that excessive contextual information may hinder the model's ability to identify and leverage relevant preference patterns.

\begin{table}[t]
\centering
\renewcommand{\arraystretch}{1}
\caption{Performance comparison of UCP, CoT, and BRP across datasets, with percentage improvements of UCP over BRP and CoT.}
\label{tab:results_comparison}
\begin{tabular}{cccccc}
\toprule
\multirow{2}{*}{Dataset} & \multirow{2}{*}{UCP} & \multirow{2}{*}{CoT} & \multirow{2}{*}{BRP} & \multicolumn{2}{c}{Improvement of UCP} \\ \cmidrule(l){5-6} 
                                 &                               &                               &                               & vs BRP      & vs CoT     \\ \midrule
MovieLens                         & 0.2460                        & 0.2240                        & 0.2000                        & +23.00\%             & +9.82\%              \\ 
LastFM                            & 0.4372                        & 0.4108                        & 0.3770                        & +15.97\%             & +6.43\%              \\ 
GoodReads                         & 0.3662                        & 0.3446                        & 0.3104                        & +17.98\%             & +6.27\%              \\ 
\bottomrule
\end{tabular}
\end{table}

\begin{figure*}[h]
    \centering
    \includegraphics[width=\textwidth]{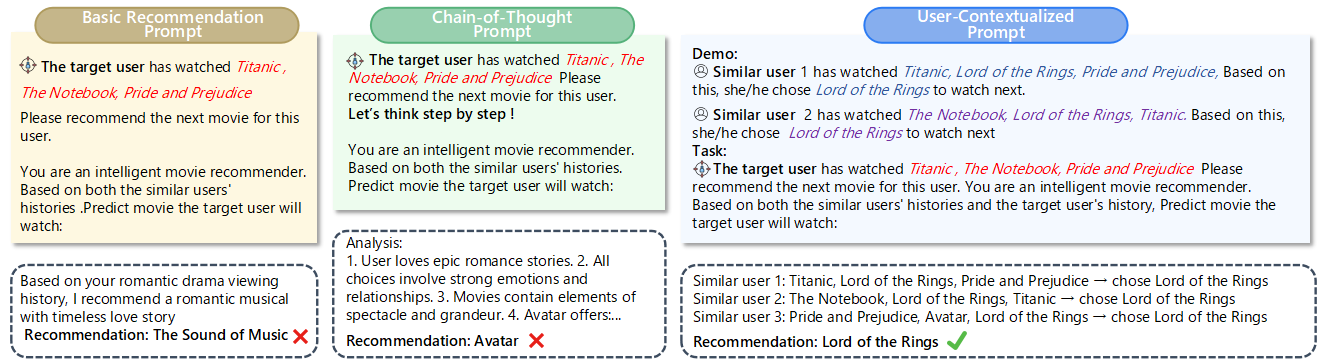}
    \caption{Illustration of User-Contextualized Prompt, Chain-of-Thought Prompt and Conventional prompt}
    \label{fig10}
\end{figure*}

\section{In-Depth Analysis}
To validate the effectiveness of our User-Contextualized Prompt (UCP) in few-shot scenarios, we conduct experiments using GPT-4 as the base model without fine-tuning. We evaluate on 100 randomly sampled test sequences from  MovieLens, LastFM, and GoodReads datasets to ensure statistical reliability, where each sequence is directly fed to GPT-4 through prompting. Then, we present a detailed case study to demonstrate how UCP's reasoning process differs from alternative prompting strategies.

\subsubsection{Few-Shot Performance}
We conduct experiments comparing UCP with two baseline prompting strategies: Chain-of-Thought (CoT) Prompts—a state-of-the-art approach that encourages step-by-step reasoning, and Basic Recommendation Prompts (BRP)—a straightforward approach using direct instruction, as illustrated in Figure~\ref{fig10}. For UCP, we select 5 similar user demonstrations based on preliminary experiments. The evaluation is performed on MovieLens, LastFM, and GoodReads datasets, with results presented in Table~\ref{tab:results_comparison}.

Across all datasets, UCP consistently demonstrates superior performance in terms of HR@1. Specifically, on MovieLens, it achieves a 23.00\% improvement over BRP and a 9.82\% improvement over CoT. Similar patterns are observed on LastFM (15.97\% over BRP, 6.43\% over CoT) and GoodReads (17.98\% over BRP, 6.27\% over CoT). These significant improvements highlight UCP's effectiveness in leveraging collaborative signals for few-shot recommendation.

\subsubsection{Qualitative Analysis}
To understand the reasoning process and recommendation effectiveness of different prompting strategies, we analyze a representative case where the target user has watched romantic dramas: Titanic, The Notebook, Pride and Prejudice. As shown in Figure \ref{fig10}, the three prompting strategies exhibit distinct recommendation behaviors:

\begin{itemize}[leftmargin=1em]
   \item\textbf{BRP} focuses solely on genre matching, recommending The Sound of Music based on its romantic elements. This recommendation proves incorrect, highlighting its limitation in considering only surface-level genre similarities.
   
   \item \textbf{CoT} demonstrates more sophisticated reasoning by analyzing multiple factors like epic scale and directorial connections, leading to recommending Avatar. Despite its logical reasoning process, this recommendation also fails to match the user's actual next watch.
   
   \item \textbf{UCP} identifies a non-intuitive but data-supported pattern through similar users' behaviors, recommending Lord of the Rings. This cross-genre recommendation proves correct as it is validated by multiple similar users' actual viewing patterns.
   
   \item This case analysis reveals UCP's unique advantage in discovering authentic viewing patterns through collaborative signals, rather than relying on predefined genre boundaries or purely logical reasoning. The results demonstrate UCP's potential in capturing natural user interest evolution in recommendation tasks.
\end{itemize}

\section{CONCLUSION}
This paper introduces AdaptRec, a novel self-adaptive framework that leverages Large Language Models (LLMs) to address fundamental challenges in sequential recommendation systems. Our key contributions—the Self-Adaptive User Selection Paradigm and User-Contextualized Recommendation Prompt Design—enable dynamic integration of collaborative information and adaptation to evolving user contexts. Extensive experiments across multiple datasets demonstrate AdaptRec's significant and consistent performance improvements over state-of-the-art methods. These results underscore the potential of combining collaborative filtering techniques with LLMs' reasoning capabilities, marking a substantial advancement in personalized recommendation systems. AdaptRec opens new research directions in adaptive, context-aware recommendations, with implications for scalability, multi-modal integration. Our work establishes a new paradigm for leveraging LLMs in recommendation tasks, paving the way for more sophisticated, user-centric recommendation systems across diverse application domains. For future work, we aim to explore integrating more diverse data sources, such as incorporating multi-modal information like images and audio to enrich the understanding of user contexts. Additionally, we plan to investigate the scalability of AdaptRec in ultra-large-scale environments and optimize its computational efficiency.

\bibliographystyle{ACM-Reference-Format}
\bibliography{main}

\end{document}